Layer-Transferred MoS$_2$/GaN PN Diodes

Edwin W. Lee II[1], Choong Hee Lee[1], Pran K. Paul[1], Lu Ma[2], William D. McCulloch[2], Sriram Krishnamoorthy[1], Yiying Wu[2], Aaron Arehart[1], Siddharth Rajan[1,3]

**Abstract**

Electrical and optical characterization of two-dimensional/three-dimensional (2D/3D) p-molybdenum disulfide/n-gallium nitride (p-MoS$_2$/n-GaN) heterojunction diodes are reported. Devices were fabricated on high-quality, large-area p-MoS$_2$ grown by chemical vapor deposition (CVD) on hexagonal sapphire substrates. The processed devices were transferred onto GaN/sapphire substrates, and the transferred films were characterized by X-ray diffraction (XRD) and atomic force microscopy (AFM). On-axis XRD spectra and surface topology obtained from AFM scans were consistent with previously grown high-quality, continuous MoS$_2$ films. Current-voltage measurements of these diodes exhibited excellent rectification, and capacitance-voltage measurements were used to extract a conduction band offset of approximately 0.2 eV for the transferred MoS$_2$/GaN heterojunction. This conduction band offset was confirmed by internal photoemission (IPE) measurements. The energy band lineup of the MoS$_2$/GaN heterojunction is proposed here. This work demonstrates the potential of 2D/3D heterojunctions for novel device applications.

Molybdenum disulfide is the most widely studied of the two-dimensional layered transition metal dichalcogenides (TMDs). Mechanically exfoliated $MoS_2$ films have been utilized to fabricate flexible and transparent devices, [1,2,3,4] photodetectors [5,6] and transistors. [7,8,9,10,11,12] Mechanical exfoliation of $MoS_2$ flakes onto bulk semiconductors circumvents the issue of lattice mismatch that might inhibit direct growth of $MoS_2$ and other 2D films onto lattice mismatched substrates. However, stable, substitutional doping of mechanically exfoliated flakes and the scalability of the device fabrication on these flakes is poor. Chemical vapor deposition (CVD) of $MoS_2$ has been reported using a variety of Mo and S precursors including $MoO_3$,[13,14,15] Mo,[16] and $MoS_2$ powder,[17] sulfur powder and $MoO_2$[18]. These growth methods result in non-continuous films, often in the form of triangular $MoS_2$ domains the largest of which are on the order of hundreds of microns. Large area films achieved by these growths rely on these domains coalescing into a polycrystalline layer.

The development of CVD growth of large-area $MoS_2$ grown on hexagonal sapphire substrates with Mo metal and S powder precursors was previously reported.[19] This growth process was improved by changing the sulfur precursor to $MoS_2$ powder and increasing the growth temperature. The resulting films were single crystal and exhibited epitaxial registry with respect to the sapphire substrate.[20] P-type conductivity in $MoS_2$ was demonstrated by adding a thin layer of Nb to the Mo metal precursor.[21]

More recently, heterojunction devices utilizing two dimensional (2D) and bulk, traditional "3D" semiconductors (SiC [22], Si [23,24,25]) have been explored. These 2D/3D heterojunctions represent an area of opportunity for the expansion of the functionality 3D semiconductors. For example, in the case of wide band gap semiconductors like GaN, SiC, and ZnO, p-type doping has proven challenging due to the high activation energy of holes. Lattice mismatch between wide and narrow band gap materials provides another constraint for integrating dissimilar semiconductors. The integration of 2D, layered, p-doped semiconductors like $MoS_2$ with wide band gap materials could provide an avenue towards achieving high-performance, bipolar devices that were otherwise unattainable due to the aforementioned obstacles.

Direct growth of high-quality, large-area, p-doped $MoS_2$ has been reported on hexagonal basal plane semiconductors like wide band gap SiC.[22] However, our previous attempts at direct growth of $MoS_2$ on GaN resulted in decomposition of GaN due to the high growth temperature and the constraints introduced by our growth methods. We combined the advantages of growth of high-quality, large-area, doped $MoS_2$ films on sapphire with the ability to transfer these films onto arbitrary substrates, allowing for the study of a wide range of 2D/3D heterostructures. We report here on the electrical and optical characterization of heterojunctions formed by thin film transfer of high-quality, CVD-grown, large-area, p-doped $MoS_2$ and n-doped GaN.

Nb-doped $MoS_2$ was grown using the growth process detailed in previous work from our group. The Mo/Nb metal precursor used to obtain degenerately doped, p-type $MoS_2$ was deposited by sputtering (AJA Orion RF/DC Sputter Deposition Tool) the metals onto c-plane sapphire substrates. A 0.2 nm layer of Nb was sandwiched between 2.5 nm layers of Mo, and the Mo/Nb-coated sapphire substrates were placed in a small quartz tube (1 cm inner diameter) with 8.0 mg of $MoS_2$ powder as a sulfur source. The tube was pumped down by mechanical pump and sealed. The samples were heated at 1100°C for 4.5 hours and then cooled at a rate of 0.5°/min as described in [Ref. 20]. The resulting film was approximately 10 nm thick, expected to exhibit a doping concentration of 3 x $10^{20}$ $cm^{-3}$ and a hole mobility of 8 $cm^2$/V s, as reported in [Ref. 21]. On-axis X-ray diffraction spectra of the transferred p-$MoS_2$ films on GaN/sapphire template are shown in figure 1a. The spectra of the transferred $MoS_2$ films showed the (002) family of peaks of 2H-$MoS_2$. This indicates that the structural quality of $MoS_2$ was not significantly degraded by the film transfer process. Figure 1b shows a 2 μm X 2 μm atomic force microscopy image of p-$MoS_2$ transferred to GaN. The root mean squared (RMS) roughness of the $MoS_2$ after transfer to GaN was 1.95 nm.

Devices were fabricated on degenerately p-doped $MoS_2$ utilizing ultra-violet stepper lithography (i-line). Ni/Au/Ni contacts were electron beam evaporated to form Ohmic contacts to $MoS_2$. Isolation mesas were dry-etched using inductively coupled plasma/reactive ion etching (ICP/RIE) with $BCl_3$/Ar chemistry. The sample was then coated with a layer of PMMA and soaked in a $NH_4OH + H_2O_2 + H_2O$ solution at 80°C.[26] Oxygen bubbles at the edges of the film detach the PMMA/$MoS_2$ film from the sapphire substrate on which it was grown and the film floats to the surface of the solution. The detached film was rinsed in DI water and fished out with a GaN substrate. GaN templates were obtained from Lumilog and were n-type doped ($N_D$=3 x $10^{18}$ $cm^{-3}$). An additional 200 nm of lightly doped n-type GaN ($N_D$=7 x $10^{17}$ $cm^{-3}$) was grown by molecular beam epitaxy on the GaN/sapphire templates to reduce band-to-band tunneling leakage in the transferred p-$MoS_2$/n-GaN diodes. Contact to the n-GaN layer was formed by indium dot (Figure 2).

Vertical current-voltage (J-V) characteristics of the $MoS_2$/GaN heterojunction diode were measured by applying bias to the Ni/Au/Ni contacts on p-$MoS_2$. The room temperature J-V characteristics showed 9 orders of magnitude rectification at +/-2 V and exhibited an ideality factor of approximately 2 in the exponential region before series resistance became dominant. This indicates that transport in the $MoS_2$/GaN heterojunction diode is likely dominated by recombination current.

Figure 4 shows the capacitance-voltage (C-V) characteristics of the $MoS_2$/GaN diode. The C-V characteristic was measured with bias applied to contacts on p-$MoS_2$ and was consistent with that of a reverse-biased p-n junction. A doping concentration of 7 x $10^{17}$ $cm^{-3}$ was extracted from the C-V measurement. This concentration was consistent with the expected doping based on secondary ion mass spectroscopy. The $1/C^2$ characteristic determined from the C-V exhibited a linear dependence on applied voltage, and the extracted built-in voltage of the $MoS_2$/GaN heterojunction was 1.5 V. We used the assumption of band gap narrowing in $MoS_2$ due to degenerate doping from [22] ($E_g$ = 950 meV), and used the Joyce-Dixon approximation to determine the position of the Fermi level in $p^+$-$MoS_2$. The conduction band offset, $\Delta E_C$, is determined by

$$\Delta E_C = qV_{bi} - E_{g,MoS_2} - \Phi_p + \Phi_n, \quad (1)$$

where $\Delta E_C$ is the conduction band offset, $E_g$ is the band gap of $MoS_2$ considering the band gap narrowing, $\Phi_p$ is the Fermi level position in p-$MoS_2$ and $\Phi_n$ is the Fermi level position in GaN. The extracted conduction band offset from the electrical characterization of the $MoS_2$/GaN heterojunction was approximately 230 meV.

Internal photoemission (IPE) measurements were used to determine the conduction band offset of the $MoS_2$/GaN heterojunction. A Xe lamp was used for optical excitation, and the resulting photocurrent was measured with an electrometer. Photocurrent was normalized to the flux variation of the lamp to rule out its contribution as the source of an onset. Zero-bias was applied to the junction, and incident photon energy was varied from 1.2 to 2.5 eV. According to [27] and Fowler's hypothesis,[28] photocurrent from the excitation of electrons from the valence band and their subsequent emission over a barrier should exhibit a quadratic dependence on photon energy. This yield corresponded to the electrons emitted per photon absorbed. Figure 5 shows the square root of the photo-yield as a function of the incident photon energy, where photo-yield, Y, is given by

$$Y = A(h\nu - E)^2 \qquad (2)$$

where A is a constant, $h\nu$ is the energy of the incident photon, and E is the total energy an electron in the valence band of $MoS_2$ requires to be excited from the valence band of $MoS_2$ to the conduction band of GaN. The extrapolation of the linear fit of the square root of photo-yield current to the energy axis shown in Figure 5 indicated a photon energy threshold of 1.46 eV. This is the energy required for the photo-excited electron to complete the process of excitation from the valence band to emission over the barrier.

We again extract the conduction band offset between $MoS_2$ and GaN using the offset determined by IPE measurements. We kept the assumption of band gap narrowing in degenerately p-doped $MoS_2$ and the calculated Fermi energy. In this case, the conduction band offset was given by

$$\Delta E_C = E_0 - (E_{g,MoS_2} + \Phi_p) \qquad (3)$$

where the $E_0$ refers to the energy threshold determined by IPE. The conduction band offset of the $MoS_2$/GaN heterojunction determined by optical measurements was approximately 170 meV.

Electrical and optical measurements of the $MoS_2$/GaN heterojunction exhibit satisfactory agreement, and the conduction band offset was found to be approximately 0.2 eV. Our work suggests that 2D/3D heterojunctions can be treated electrically using theory developed for traditional heterojunctions, and no Fermi level pinning or other effects are evident, and more significantly, the interface between $MoS_2$ and GaN showed behavior similar to traditional 3D/3D heterojunctions even though no out-of-plane bonds are formed. Figure 6 shows the band diagram of MoS$_2$/GaN. The characterization of this junction illustrates the utility of exploring 2D/3D heterojunctions to elucidate their utility for extending the functionality of existing semiconductors. MoS$_2$/GaN is appropriate as a base/collector diode in a heterojunction bipolar transistor.

We reported the electrical and optical characterization of MoS$_2$/GaN diodes formed by film transfer. The diodes exhibited excellent rectification and were ideal at room temperature. Electrical characterization of the junction by C-V measurement and optical characterization by internal photoemission yielded a conduction band offset between the materials of approximately 0.2 eV.

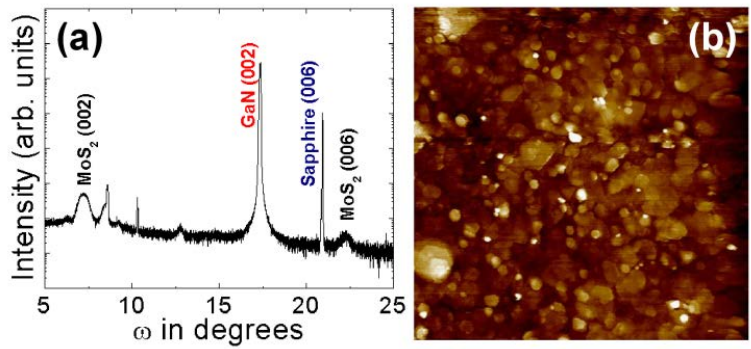

Figure 1: (a) On-axis X-ray diffraction spectra of p-doped $MoS_2$ film transferred onto GaN on sapphire templates. $MoS_2$ maintains (002) family of diffraction peaks after transfer. (b) 2 μm X 2 μm atomic force microscopy image of $MoS_2$ transferres to GaN with RMS roughness of 1.95 nm and height scale of 15 nm.

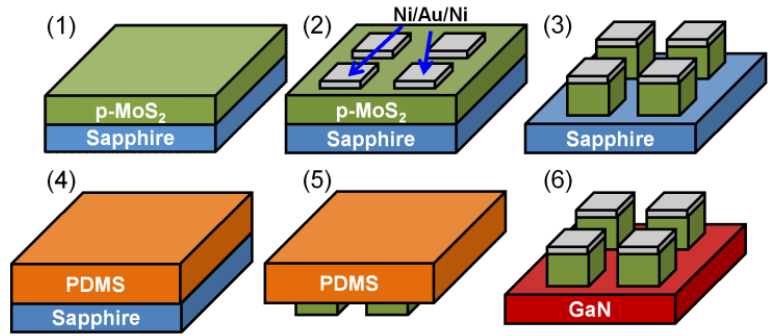

Figure 2: Transfer process for MoS$_2$/GaN diodes: (1) p-MoS$_2$ was grown on hexagonal sapphire. (2) Ohmic contacts were evaporated on MoS$_2$ and (3) devices were isolated by reactive ion etching. (4) PDMS was spun onto the processed samples and (5) PDMS-coated mesas were removed from the sapphire substrate by bubbling with NH$_4$OH + H$_2$O$_2$ + H$_2$O solution and (6) transferred to GaN. PDMS was removed, revealing the processed diodes.

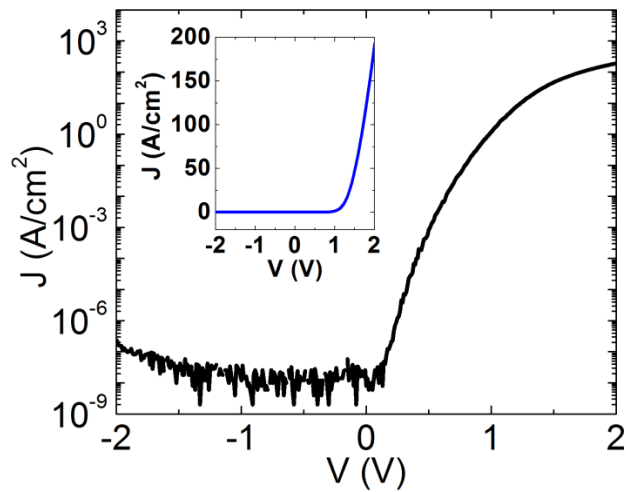

Figure 3: The current-voltage characteristic of p$^+$-MoS$_2$/n-GaN diode exhibits 9 orders of magnitude rectification. The ideality factor extracted from the room temperature characteristic is approximately 2, indicating the forward biased transport is dominated by recombination current.

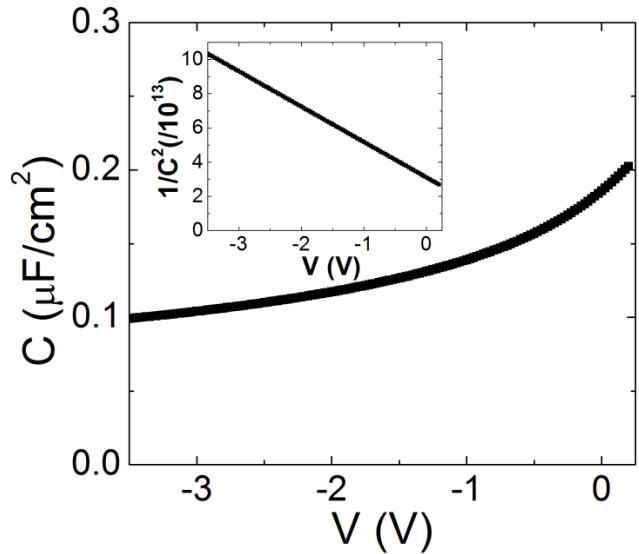

Figure 4: Capacitance-voltage characteristic of the $p^+$-$MoS_2$/n-GaN diode. The $1/C^2$ characteristic is linear with respect to voltage, and the extrapolated intercept with the voltage axis gives a built-in potential of approximately 1.5 V.

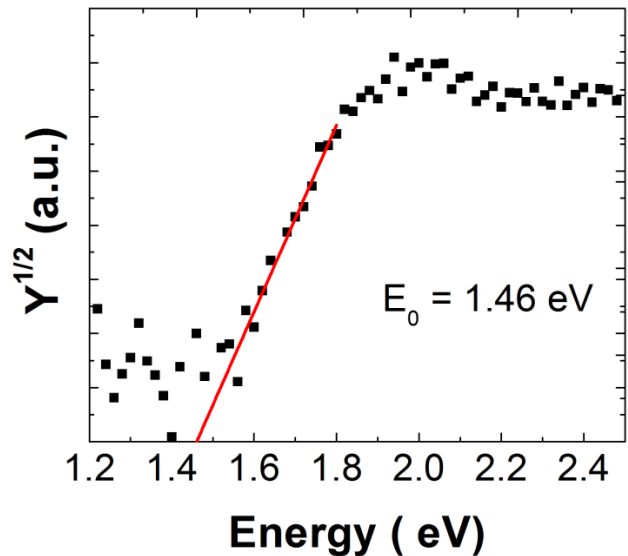

Figure 5: Internal photoemission (IPE) measurements were used to measure the barrier height between $MoS_2$ and GaN by photoexciting electrons from the valence band of $MoS_2$ and measuring the photon energy at which the carriers are able to move into GaN. This onset of 1.46 eV corresponds with a conduction band offset of 0.2 eV when band gap narrowing due to degenerate doping and the corresponding Fermi level position were accounted for.

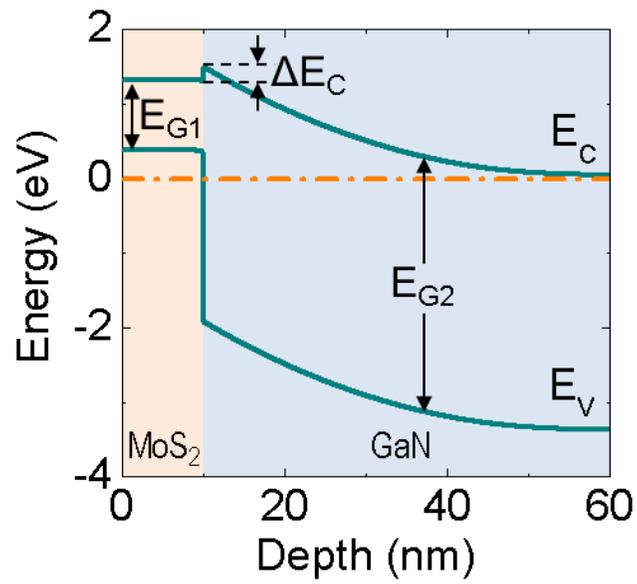

Figure 6: Band diagram of MoS$_2$/GaN. The conduction band offset of approximately 0.2 eV is determined by both optical and electrical characterization of the junction. The Fermi level (E$_F$) is shown at zero energy. E$_{G1}$ and E$_{G2}$ refer to the band gaps of MoS$_2$ and GaN, respectively.